\documentstyle{article}

\newcommand{\ba}{\begin{eqnarray}}
\newcommand{\ea}{\end{eqnarray}}
\input{psfig}

\begin{document}
\pagestyle{plain}

\title{Algebraic treatment of three-body problems}
\author{R. Bijker\\
Instituto de Ciencias Nucleares, \\
Universidad Nacional Aut\'onoma de M\'exico, \\
A.P. 70-543, 04510 M\'exico, D.F., M\'exico
\and
A. Leviatan\\
Racah Institute of Physics, The Hebrew University,\\
Jerusalem 91904, Israel}
\date{}
\maketitle

\begin{abstract}
We discuss an algebraic treatment of three-body systems 
in terms of a $U(7)$ spectrum generating algebra. 
In particular, we develop the formalism for nonlinear 
configurations and present an algebraic description 
of vibrational and rotational excitations of symmetric 
(X$_3$) and asymmetric tops (XY$_2$ and XYZ). The relevant 
point group symmetry is incorporated exactly. 
\end{abstract}

\begin{center}
\end{center}

\clearpage

\section{Introduction}

The study of few-body problems has played an important role
in many fields of physics and chemistry \cite{fbs}.
Over the years accurate methods have
been developed to solve the few-body equations. The degree of
sophistication required depends on the physical system, {\it i.e.}
to solve the few-body problem in atomic physics requires a far
higher accuracy than in hadronic physics.

In recent years, the development and application of algebraic methods
to the many-body problem ({\it e.g.} collective excitations in nuclei
\cite{ibm} and molecules \cite{vibron}) has received considerable
attention. In spectroscopic studies these algebraic methods
provide a powerful tool to study symmetries and selection
rules, to classify the basis states, and to calculate matrix elements.
In this paper we discuss the application of algebraic methods to the
few-body problem. Especially in the area of hadronic physics, which
is that of strong interactions at low energies, for which exact
solutions of QCD are unavailable, these methods may become very useful.
The approach is based on the 
general criterion \cite{FI} to consider $U(k+1)$ as a spectrum generating
algebra for a bound-state problem with $k$ degrees of freedom, and
assigning all states to the symmetric representation $[N]$ of $U(k+1)$.
For the $k=5$ quadrupole degrees of freedom in 
collective nuclei this led to the introduction of the $U(6)$ 
interacting boson model \cite{ibm} and for the $k=3$ dipole 
degrees of freedom in diatomic molecules to the 
$U(4)$ vibron model \cite{vibron}. 

In a three-body system the dynamics is determined by the six degrees
of freedom of two relative vectors, which in the algebraic
approach leads to a $U(7)$ spectrum generating algebra. 
This model was originally introduced for a 
system of three identical objects \cite{BIL,BDL}. 
In this paper, we develop the general formalism for a system of three
objects and discuss, in particular, an algebraic description
of vibrational and rotational excitations of symmetric and
asymmetric tops. Applications can be found in molecular physics 
(XYZ, XY$_2$ and X$_3$ molecules) and hadronic physics ($qqq$ baryons). 

\section{Algebraic treatment of a three-body system}

The internal motion of a three-body system can be described in
terms of the relative Jacobi coordinates, 
$\vec{\rho}$ and $\vec{\lambda}$, which we choose as 
\ba
\vec{\rho} &=& \frac{1}{\sqrt{2}} \, (\vec{r}_1 - \vec{r}_2) ~, 
\nonumber\\
\vec{\lambda} &=& 
\frac{1}{\sqrt{m_1^2+m_2^2+(m_1+m_2)^2}} \,  
[ m_1 \vec{r}_1 + m_2 \vec{r}_2 - (m_1+m_2) \vec{r}_3 ] ~. 
\label{jacobi}
\ea
Here $m_i$ and $\vec{r}_i$ denote the mass and coordinate of the 
$i$-th object. For three identical objects with equal masses,  
Eq.~(\ref{jacobi}) reduces to the Jacobi coordinates used in \cite{KM}.
Instead of a formulation in terms of coordinates and momenta, we 
use the method of bosonic quantization in which we introduce a dipole 
boson with angular momentum and parity $L^P=1^-$ for each independent 
relative coordinate, and an auxiliary scalar boson with $L^P=0^+$ 
\ba
p^{\dagger}_{\rho,m} ~, \; p^{\dagger}_{\lambda,m} ~, \;
s^{\dagger} \hspace{1cm} (m=-1,0,1) ~. \label{bb}
\ea
The scalar boson does not represent an independent degree of freedom,
but is added under the restriction that the total number of bosons
$N=n_{\rho}+n_{\lambda}+n_s$ is conserved. This procedure leads
to a spectrum generating algebra of $U(7)$ \cite{BIL,BDL}. 
For a system of interacting bosons the model space is spanned by the
symmetric irreducible representation $[N]$ of $U(7)$, which contains
the harmonic oscillator shells with $n=n_{\rho}+n_{\lambda}=0,1,\ldots,N$.
The value of $N$ determines the size of the model space.

In case two or all three objects are identical, the Hamiltonian 
has to be invariant under the permutation group, 
$S_2$ or $S_3$, respectively. In the former case, in which we label 
the identical objects by 1 and 2, the permutation symmetry is determined
by the transposition $P(12)$, and in the latter case by both $P(12)$
and the cyclic permutation $P(123)$ \cite{KM}.  
All other permutations can be expressed in terms of these two 
elementary ones. Algebraically, these operators can be transcribed as
\ba
P(12) &=& \mbox{exp} \left\{ -i \, \pi \hat n_{\rho} \right\} ~,
\nonumber\\
P(123) &=& \mbox{exp} \left\{ -i \, \frac{2\pi}{3} \hat F_2 \right\} ~,
\label{point}
\ea
with
\ba
\hat n_{\rho} &=& \sum_{m} p^{\dagger}_{\rho,m} p_{\rho,m} ~,
\nonumber\\
\hat F_2 &=& -i \sum_{m} \left( p^{\dagger}_{\rho,m} p_{\lambda,m}
- p^{\dagger}_{\lambda,m} p_{\rho,m}  \right) ~. \label{fspin}
\ea
For three identical objects the eigenstates of the Hamiltonian are 
characterized by the irreducible representations of the $S_3$ 
permutation group \cite{KM,Mitra}. 
However, in anticipation of the geometric analysis of the next 
sections, we use a labeling under the point group $D_3$ (which is
isomorphic to $S_3$): namely $A_1$ and $A_2$ for the one-dimensional
symmetric and antisymmetric representations, and $E$ for the 
two-dimensional mixed symmetry representation.
The scalar boson $s^{\dagger}$ of Eq.~(\ref{bb}) transforms as the
symmetric representation $A_1$, whereas the two dipole bosons
$p^{\dagger}_{\rho,m}$ and $p^{\dagger}_{\lambda,m}$ transform as
the two components of the mixed symmetry representation,
$E_{\rho}$ and $E_{\lambda}$, respectively.
For two identical objects the Hamiltonian is invariant under 
$S_2$ (which is isomorphic to $D_2$), and the $D_3$ irreducible 
representations reduce to those of its $D_2$ subgroup
as $A_1,E_{\lambda}\rightarrow A$ (symmetric) 
and $A_2,E_{\rho}\rightarrow B$ (antisymmetric).

All physical operators, such as the Hamiltonian and transition 
operators, are expressed in terms of the building blocks 
of Eq.~(\ref{bb}). With the help of the transposition $P(12)$ and the
cyclic permutation $P(123)$ we construct a set of generators of 
the algebra of $U(7)$ that transform as tensor operators under the 
rotation group $SO(3)$ and the point group chain $D_3 \supset D_2$ 
\ba
\begin{array}{llll}
\hat D_{\rho,m} \;=\; (p^{\dagger}_{\rho} s -
s^{\dagger} \tilde{p}_{\rho})^{(1)}_m & (E_{\rho},B) ~, & 
\hspace{1cm} \hat D_{\lambda,m} \;=\; (p^{\dagger}_{\lambda} s -
s^{\dagger} \tilde{p}_{\lambda})^{(1)}_m & (E_{\lambda},A) ~, \\
\hat A_{\rho,m} \;=\; i \, (p^{\dagger}_{\rho} s +
s^{\dagger} \tilde{p}_{\rho})^{(1)}_m & (E_{\rho},B) ~, & 
\hspace{1cm} 
\hat A_{\lambda,m} \;=\; i \, (p^{\dagger}_{\lambda} s +
s^{\dagger} \tilde{p}_{\lambda})^{(1)}_m & (E_{\lambda},A) ~, \\
\hat G^{(l)}_{\rho,m} \;=\; ( p^{\dagger}_{\rho} \tilde{p}_{\lambda}
+ p^{\dagger}_{\lambda} \tilde{p}_{\rho} )^{(l)}_m & (E_{\rho},B) ~, & 
\hspace{1cm} 
\hat G^{(l)}_{\lambda,m} \;=\; ( p^{\dagger}_{\rho} \tilde{p}_{\rho}
- p^{\dagger}_{\lambda} \tilde{p}_{\lambda} )^{(l)}_m & (E_{\lambda},A) ~, \\
\hat G^{(l)}_{A_1,m} \;=\; ( p^{\dagger}_{\rho} \tilde{p}_{\rho}
+ p^{\dagger}_{\lambda} \tilde{p}_{\lambda} )^{(l)}_m & (A_1,A) ~, & 
\hspace{1cm}
\hat G^{(l)}_{A_2,m} \;=\; i \, ( p^{\dagger}_{\rho} \tilde{p}_{\lambda}
- p^{\dagger}_{\lambda} \tilde{p}_{\rho} )^{(l)}_m & (A_2,B) ~, \\
\hat{n}_s \;=\; s^{\dagger}s & (A_1,A) ~, & & 
\end{array} 
\label{gen}
\ea
with $l=0,1,2$. Here $\tilde{p}_{\rho,m} = (-1)^{1-m} p_{\rho,-m}$ and 
$\tilde{p}_{\lambda,m} = (-1)^{1-m} p_{\lambda,-m}$.
The $U(7)$ Hamiltonian can be expressed in terms of scalar 
products of the operators of Eq.~(\ref{gen}). 
The eigenvalues and corresponding eigenvectors can be obtained exactly
by diagonalization in an appropriate basis. 

\section{Geometry and intrinsic-collective structure}

A more intuitive geometric interpretation of algebraic Hamiltonians 
can be obtained by using mean-field techniques. The main ingredient 
is the introduction of coherent (or intrinsic) states 
as variational wave functions which for 
a system of bosons have the form of a condensate of $N$ deformed 
bosons \cite{cs}. In the present case of $U(7)$, the condensate 
can be parametrized as 
\ba
| N;r,\chi,\theta \rangle &=&
\frac{1}{\sqrt{N!}} \, (b^{\dagger}_c)^N \, | 0 \rangle ~,
\label{cond}
\ea
with
\ba
b_c^{\dagger} &=& \left[ s^{\dagger}
+ r \cos \chi \, p_{\lambda,x}^{\dagger}
+ r \sin \chi \, (\cos \theta \, p_{\rho,x}^{\dagger} 
+  \sin \theta \, p_{\rho,y}^{\dagger}) \right]/\sqrt{1+r^2} ~.
\label{bc}
\ea
The geometry is chosen such that the $xy$ plane is defined by 
$\vec{\rho}$ and $\vec{\lambda}$, with the $x$-axis along 
$\vec{\lambda}$ and the $z$-axis perpendicular to this plane. 
The two vectors $\vec{\rho}$ and $\vec{\lambda}$ 
are parametrized in terms of the three Euler angles which 
are associated with the orientation of the three-body system, 
and three internal coordinates which are taken as the two lengths 
of the vectors $r_{\rho}$ and $r_{\lambda}$, and their relative 
angle $\theta$: $\vec{r}_{\lambda} \cdot \vec{r}_{\rho} 
= r_{\lambda} r_{\rho} \cos \theta$~.
The two lengths are parametrized in terms of the hyperspherical 
radius $r$ and the hyperangle $\chi$: $r_{\lambda} = r \cos \chi$ and 
$r_{\rho} = r \sin \chi$~. The hyperradius $r$ is a measure of the 
size of the system, whereas the hyperangle 
$\chi$ and the angle $\theta$ determine its shape \cite{hyper}.
Since the Hamiltonian is rotationally invariant, there is no need 
to introduce the Euler angles in Eq.~(\ref{bc}).
The expectation value of the Hamiltonian 
in the coherent state defines its energy surface
\ba
E_{N}(r,\chi,\theta) &=& \langle N;r,\chi,\theta | \, H \,
| N;r,\chi,\theta \rangle ~.
\ea 
The minimum of the energy surface determines the equilibrium values 
($\overline{r},\overline{\chi},\overline{\theta}$) of the geometric 
variables. The corresponding condensate wave function 
$|N;\overline{r},\overline{\chi},\overline{\theta} \rangle$  
represents the equilibrium shape of the three-body system. 

In order to analyze the vibrational and rotational excitations, 
it is convenient to split the Hamiltonian into an intrinsic
(vibrational) and a collective (rotational) part \cite{KL}
\ba
H &=& H_{\mbox{int}} + H_{\mbox{c}} ~. \label{ham}
\ea
The intrinsic part, by definition, annihilates the
equilibrium condensate and has, up to an overall constant, the same
energy surface as the original Hamiltonian
\ba
H_{\mbox{int}} \, | N;\overline{r},\overline{\chi},
\overline{\theta} \rangle &=& 0 ~,
\nonumber\\
\langle N;r,\chi,\theta | \, H_{\mbox{int}} \,
| N;r,\chi,\theta \rangle &=&
E_{N}(r,\chi,\theta) - E_0 ~.
\ea
For $\overline{r}>0$ the rotational symmetry is spontaneously broken. 
Although in this case the condensate of Eq.~(\ref{cond}) is deformed,
it is still an eigenstate of the rotationally invariant intrinsic 
Hamiltonian. Since $H$ and $H_{\mbox{int}}$ have
the same energy surface, one can extract the normal modes of the
system by carrying out a normal mode analysis on $H_{\mbox{int}}$. 
The collective part is the remainder of the Hamiltonian
$H_{\mbox{c}} = H - H_{\mbox{int}}$.

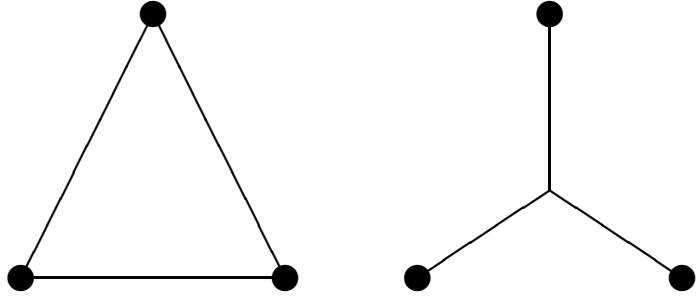
\begin{figure}
\centering
\setlength{\unitlength}{1.0pt}
\begin{picture}(300,200)(0,0)
\thicklines
\put ( 25, 50) {\circle*{10}}
\put (125, 50) {\circle*{10}}
\put ( 75,150) {\circle*{10}}
\put ( 25, 50) {\line ( 1,0){100}}
\put ( 25, 50) {\line ( 1,2){ 50}}
\put (125, 50) {\line (-1,2){ 50}}

\put (175, 50) {\circle*{10}}
\put (275, 50) {\circle*{10}}
\put (225,150) {\circle*{10}}
\put (175, 50) {\line ( 3, 2){ 50}}
\put (275, 50) {\line (-3, 2){ 50}}
\put (225,150) {\line ( 0,-1){ 67}}
\end{picture}
\caption[]{\small Geometry of a three-body system.
\normalsize} \label{geometry}
\end{figure}

Rather than starting from the most general algebraic Hamiltonian, 
analyzing its energy surface, determining its equilibrium shapes, 
and decomposing the Hamiltonian for each one of them into an 
intrinsic and collective part \cite{KL}, here we start from a given 
equilibrium shape and construct the corresponding intrinsic 
and collective Hamiltonians to analyze the vibrational and 
rotational excitations. In this paper we discuss the nonlinear 
equilibrium shapes of three objects, see Fig.~\ref{geometry}. 

\subsection{Intrinsic Hamiltonian}

Nonlinear configurations are characterized by the equilibrium values 
of the geometric variables 
\ba
\overline{r} \;=\; R \;>\; 0 ~, 
\hspace{1cm} \overline{\chi} \;=\; \beta ~, 
\hspace{1cm} \overline{\theta} \;=\; \gamma \;\neq\; 0,\pi ~. 
\label{xyz}
\ea
The corresponding intrinsic Hamiltonian can be expressed in the form
\ba 
H_{\mbox{int}} &=& A \, P_1^{\dagger} P_1 + B \, P_2^{\dagger} P_2 
+ C \, P_3^{\dagger} P_3 
\nonumber\\
&& + D \, ( P_1^{\dagger} P_2 + P_2^{\dagger} P_1 ) 
   + E \, ( P_1^{\dagger} P_3 + P_3^{\dagger} P_1 ) 
   + F \, ( P_2^{\dagger} P_3 + P_3^{\dagger} P_2 ) ~, 
\label{hint1} 
\ea 
with 
\ba 
P_1^{\dagger} &=& p^{\dagger}_{\rho} \cdot p^{\dagger}_{\rho}
+ p^{\dagger}_{\lambda} \cdot p^{\dagger}_{\lambda}
- R^2 \, s^{\dagger} s^{\dagger} ~,
\nonumber\\
P_2^{\dagger} &=& 
  \sin^2 \beta \, p^{\dagger}_{\lambda} \cdot p^{\dagger}_{\lambda}
- \cos^2 \beta \, p^{\dagger}_{\rho} \cdot p^{\dagger}_{\rho} ~,
\nonumber\\
P_3^{\dagger} &=&
\sin (2\beta) \, p^{\dagger}_{\rho} \cdot p^{\dagger}_{\lambda}
- \cos \gamma \, \left (\, \sin^2 \beta \,
p^{\dagger}_{\lambda} \cdot p^{\dagger}_{\lambda}
+ \cos^2 \beta \, p^{\dagger}_{\rho} \cdot p^{\dagger}_{\rho} \, \right ) ~. 
\label{hint2}
\ea

\subsection{Collective Hamiltonian}

By construction, the collective Hamiltonian has a completely flat 
(or structureless) energy surface, which does not depend on the 
geometric variables ($r,\chi,\theta$). 
Its most general form can be found by examining the structure
of the condensate boson which defines the energy surface.
The condensate boson of Eq.~(\ref{bc}) is related to the scalar
boson $s^{\dagger}$ by a unitary transformation 
\ba
b_c^{\dagger} &=& U(r,\chi,\theta) \, s^{\dagger} \, 
U^{-1}(r,\chi,\theta) ~,
\nonumber\\
U(r,\chi,\theta) &=& \mbox{e}^{-i \theta \hat L_{\rho,z}} \,
\mbox{e}^{i \chi \hat F_2} \, \mbox{e}^{-i \alpha \hat A_{\lambda,x}} ~, 
\label{trans} 
\ea
with $\cos \alpha = 1/\sqrt{1+r^2}$ and $\sin \alpha = r/\sqrt{1+r^2}$~. 
The operators $\hat L_{\rho,m}$ and $\hat F_2$ can be expressed in 
terms of the generators of Eq.~(\ref{gen}) as 
$\hat L_{\rho,m} = ( \hat G^{(1)}_{A_1,m} 
+ \hat G^{(1)}_{\lambda,m} )/\sqrt{2}$ and 
$\hat F_2 = - \sqrt{3} \, \hat G^{(0)}_{A_2}$~. 
The generators of the transformation of Eq.~(\ref{trans}) all belong 
to the $\overline{SO(7)}$ subgroup of $U(7)$, which is generated by
\ba 
\hat A_{\rho,m} ~, \; \hat A_{\lambda,m} ~, \;
\hat G^{(1)}_{\rho,m} ~, \; \hat G^{(1)}_{\lambda,m} ~, \;
\hat G^{(1)}_{A_1,m} ~, \;
\hat G^{(0)}_{A_2,0} ~, \; \hat G^{(2)}_{A_2,m} ~. \label{so7}
\ea
For this reason, the collective Hamiltonian
can be expressed in terms of the two-body part of scalar products of
these generators and has the following $D_3$ decomposition
\ba
H_{\mbox{c}} &=& H^{(A_1)}_{\mbox{c}} 
+ H^{(E_{\lambda})}_{\mbox{c}} + H^{(E_{\rho})}_{\mbox{c}} ~, 
\label{hcol1} 
\ea
where
\ba
H^{(A_1)}_{\mbox{c}}  &=& \kappa_1 \, : \left[
\hat A_{\rho} \cdot \hat A_{\rho} + \hat A_{\lambda} \cdot
\hat A_{\lambda} \right] :
+ 2\kappa_2 \, : \left[ \hat G^{(1)}_{\rho} \cdot \hat G^{(1)}_{\rho}
+ \hat G^{(1)}_{\lambda} \cdot \hat G^{(1)}_{\lambda} \right] :
\nonumber\\
&& + \, \kappa_3 \, : \hat L^{(1)} \cdot \hat L^{(1)} :
+ \kappa_4 \, : \hat F_2^2 : ~,
\nonumber\\
H^{(E_{\lambda})}_{\mbox{c}}  &=& \kappa_1' \, : \left[
\hat A_{\rho} \cdot \hat A_{\rho} - \hat A_{\lambda} \cdot
\hat A_{\lambda} \right] :
+ 2\kappa_2' \, : \left[ \hat G^{(1)}_{\rho} \cdot \hat G^{(1)}_{\rho}
- \hat G^{(1)}_{\lambda} \cdot \hat G^{(1)}_{\lambda} \right] :
\nonumber\\
&& + \, \kappa_3' \sqrt{2} \, : \left [ \hat L^{(1)}
\cdot \hat G^{(1)}_{\lambda} +
\hat G^{(1)}_{\lambda}\cdot \hat L^{(1)} \right ] : ~,
\nonumber\\
H^{(E_{\rho})}_{\mbox{c}}  &=& \kappa_1'' \, : \left[
\hat A_{\rho} \cdot \hat A_{\lambda} + \hat A_{\lambda} \cdot
\hat A_{\rho} \right] :
+ \kappa_2'' \, : \left[ \hat G^{(1)}_{\rho} \cdot \hat G^{(1)}_{\lambda}
+ \hat G^{(1)}_{\lambda} \cdot \hat G^{(1)}_{\rho} \right] :
\nonumber\\
&& + \, \kappa_3'' \sqrt{2} \, : \left [ \hat L^{(1)}
\cdot \hat G^{(1)}_{\rho} +
\hat G^{(1)}_{\rho}\cdot \hat L^{(1)} \right ] : ~.
\label{hcol2}
\ea
Here $\hat L^{(1)}_m  = \sqrt{2} \, \hat G^{(1)}_{A_1,m}$ represents the
angular momentum operator, and $::$ denotes normal ordering. 
Note that the $A_1$ term $G^{(2)}_{A_2}\cdot G^{(2)}_{A_2}$ does
not appear in Eq. (\ref{hcol2}) since it is not independent
\ba
\hat G^{(2)}_{A_2}\cdot \hat G^{(2)}_{A_2} &=&
\hat G^{(1)}_{\rho}\cdot \hat G^{(1)}_{\rho} +
\hat G^{(1)}_{\lambda}\cdot \hat G^{(1)}_{\lambda}
- \hat G^{(1)}_{A_1}\cdot \hat G^{(1)}_{A_1} +
2 \hat G^{(0)}_{A_2}\cdot \hat G^{(0)}_{A_2} ~.
\ea
The collective Hamiltonian shifts, splits and
generally mixes the bands generated by $H_{\mbox{int}}$.

\section{Asymmetric tops}

We first discuss nonlinear XYZ and XY$_2$ configurations which 
correspond to asymmetric tops. 

\subsection{Nonlinear XYZ configurations}

For a system of three different objects XYZ, there are no 
restrictions arising from the $C_s$ point group symmetry 
\cite{Herzberg}. 
The equilibrium shape is given by Eq.~(\ref{xyz}), 
and hence all terms in the Hamiltonian of Eq.~(\ref{ham}) 
are allowed. The intrinsic part is given by Eqs.~(\ref{hint1}) 
and (\ref{hint2}), and the collective part by Eqs.~(\ref{hcol1}) 
and (\ref{hcol2}). 
The normal modes can be obtained by carrying out a 
normal mode analysis on $H_{\mbox{int}}$. This is done by expressing 
it in terms of a deformed boson basis, which is spanned by the 
condensate boson of Eq.~(\ref{bc}) with 
$r=R$, $\chi=\beta$ and $\theta=\gamma$ ($\neq 0$, $\pi$) 
\ba
b_c^{\dagger} &=& \left[ s^{\dagger}
+ R \cos \beta \, p_{\lambda,x}^{\dagger}
+ R \sin \beta \, (\cos \gamma \, p_{\rho,x}^{\dagger} 
+  \sin \gamma \, p_{\rho,y}^{\dagger}) \right]/\sqrt{1+R^2} ~,
\label{b0}
\ea
and six additional orthonormal fluctuation bosons
\ba
b_u^{\dagger} &=& \left[ -R \, s^{\dagger}
+ \cos \beta \, p_{\lambda,x}^{\dagger}
+ \sin \beta \, (\cos \gamma \, p_{\rho,x}^{\dagger} 
+  \sin \gamma \, p_{\rho,y}^{\dagger}) \right]/\sqrt{1+R^2} ~,
\nonumber\\
b^{\dagger}_{v} &=& \sin \beta \, p_{\lambda,x}^{\dagger}
- \cos \beta \, (\cos \gamma \, p^{\dagger}_{\rho,x}
+ \sin \gamma \, p_{\rho,y}^{\dagger}) ~,
\nonumber\\
b^{\dagger}_{w} &=& \sin \beta \, p_{\lambda,y}^{\dagger}
+ \cos \beta \, (\sin \gamma \, p^{\dagger}_{\rho,x}
- \cos \gamma \, p_{\rho,y}^{\dagger}) ~,
\nonumber\\
b^{\dagger}_{1} &=& p_{\rho,z}^{\dagger} ~,
\nonumber\\
b^{\dagger}_{2} &=& p_{\lambda,z}^{\dagger} ~,
\nonumber\\
b^{\dagger}_{3} &=& \cos \beta \, p_{\lambda,y}^{\dagger}
- \sin \beta \, (\sin \gamma \, p^{\dagger}_{\rho,x}
- \cos \gamma \, p_{\rho,y}^{\dagger}) ~.
\label{bint}
\ea
The mean-field Hamiltonian $^{B}H_{\mbox{int}}$ 
is obtained by the usual Bogoliubov treatment of bosonic systems 
which amounts to replacing the condensate boson operators 
$b^{\dagger}_c$ and $b_c$ by $\sqrt{N}$ and keeping only terms to 
leading order in $N$ \cite{KL}
\ba
^{B}H_{\mbox{int}} 
&=& \epsilon_u \, b^{\dagger}_{u} b_{u}
+ \epsilon_v \, b^{\dagger}_{v} b_{v}
+ \epsilon_w \, b^{\dagger}_{w} b_{w}
+ \epsilon_{uv} (b^{\dagger}_{u} b_{v} + b^{\dagger}_{v} b_{u}) 
\nonumber\\
&& + \epsilon_{uw} (b^{\dagger}_{u} b_{w} + b^{\dagger}_{w} b_{u})
   + \epsilon_{vw} (b^{\dagger}_{v} b_{w} + b^{\dagger}_{w} b_{v}) ~,
\ea
with
\ba
\epsilon_u &=& 4ANR^2 ~,
\nonumber\\
\epsilon_v &=& BNR^2 \sin^2 (2\beta)/(1+R^2) ~,
\nonumber\\
\epsilon_w &=& CNR^2 \sin^2 (2\beta) \sin^2 \gamma/(1+R^2) ~,
\nonumber\\
\epsilon_{uv} &=& 2DNR^2 \sin (2\beta)/\sqrt{1+R^2} ~,
\nonumber\\
\epsilon_{uw} &=& 2ENR^2 \sin (2\beta) \sin \gamma/\sqrt{1+R^2} ~,
\nonumber\\
\epsilon_{vw} &=& FNR^2 \sin^2 (2\beta) \sin \gamma/(1+R^2) ~.
\ea
The spontaneously broken rotational symmetry ($R>0$) ensures that 
only the vibrational bosons $b_i^{\dagger}$ ($i=u,v,w$)  
show up in $^{B}H_{\mbox{int}}$. 
The intrinsic modes of the system which correspond to small 
oscillations about the minimum of the energy surface can be 
obtained by diagonalizing $^{B}H_{\mbox{int}}$. The bosons 
$b_i^{\dagger}$ ($i=1,2,3$) are associated with rotations 
of the equilibrium condensate, and hence can be identified with the 
Goldstone modes of the spontaneously broken rotational symmetry.

The mean-field or Bogoliubov image of the collective 
Hamiltonian has the form 
\ba
^{B}H_{\mbox{c}} 
&=&  \eta_u \, ( b^{\dagger}_u - b_u )^2 
   + \eta_v \, ( b^{\dagger}_v - b_v )^2 
   + \eta_w \, ( b^{\dagger}_w - b_w )^2 
\nonumber\\
&& + \eta_1 \, ( b^{\dagger}_1 - b_1 )^2 
   + \eta_2 \, ( b^{\dagger}_2 - b_2 )^2 
   + \eta_3 \, ( b^{\dagger}_3 - b_3 )^2 
\nonumber\\
&& + \eta_{uv} \, ( b^{\dagger}_u - b_u ) \, ( b^{\dagger}_v - b_v )
   + \eta_{uw} \, ( b^{\dagger}_u - b_u ) \, ( b^{\dagger}_w - b_w )  
   + \eta_{vw} \, ( b^{\dagger}_v - b_v ) \, ( b^{\dagger}_w - b_w )
\nonumber\\
&& + \eta_{12} \, ( b^{\dagger}_1 - b_1 ) \, ( b^{\dagger}_2 - b_2 )
   + \eta_{3v} \, ( b^{\dagger}_3 - b_3 ) \, ( b^{\dagger}_v - b_v )
   + \eta_{3w} \, ( b^{\dagger}_3 - b_3 ) \, ( b^{\dagger}_w - b_w ) ~,
\ea
with
\ba
\eta_u &=& -N \left\{ \kappa_1 - \kappa_1' \cos(2\beta) 
+ \kappa_1'' \cos \gamma \sin(2\beta) \right\} ~,
\nonumber\\
\eta_v &=& -N \left\{ \kappa_1 + R^2 
\left[ \kappa_2 \sin^2 \gamma 
+ \kappa_4 \cos^2 \gamma \right] \right\} /(1+R^2) 
\nonumber\\
&& -N \left\{ \kappa_1' \cos(2\beta) + R^2 \kappa_2' \sin^2 \gamma 
\right\} /(1+R^2) 
\nonumber\\
&& +N \kappa_1'' \cos \gamma \sin(2\beta) /(1+R^2) ~,
\nonumber\\
\eta_w &=& -N \left\{ \kappa_1 + R^2 
\left[ \kappa_2 
\left( \cos^2 \gamma \cos^2(2\beta)+\sin^2(2\beta) \right)
+ \kappa_4 \sin^2 \gamma \cos^2(2\beta) \right] \right\} /(1+R^2)
\nonumber\\
&& -N \left\{ \kappa_1' \cos(2\beta) + R^2 \kappa_2' 
\left[ -\sin^2(2\beta)+\cos^2 \gamma \cos^2(2\beta) \right] 
\right\} /(1+R^2) 
\nonumber\\
&& +N \left\{ \kappa_1'' - R^2 \kappa_2'' \cos(2\beta) \right\}
\cos \gamma \sin(2\beta)/(1+R^2) ~, 
\ea
and
\ba 
\eta_1 &=& -N \left\{ \kappa_1 + R^2 
\left[ \kappa_2 + \kappa_3 \sin^2 \beta \right] 
\right\} /(1+R^2) 
\nonumber\\
&& -N \left\{ \kappa_1' + R^2 \left[ \kappa_2' \cos(2\beta) 
+ 2\kappa_3' \sin^2 \beta \right] \right\} /(1+R^2) 
\nonumber\\
&& -NR^2 \left[ \kappa_2'' + 2\kappa_3'' \right] 
\cos \gamma \sin \beta \cos \beta /(1+R^2) ~,
\nonumber\\
\eta_2 &=& -N \left\{ \kappa_1 + R^2 
\left[ \kappa_2 + \kappa_3 \cos^2 \beta \right] 
\right\} /(1+R^2)
\nonumber\\
&& -N \left\{ -\kappa_1' + R^2 \left[ -\kappa_2' \cos(2\beta) 
- 2\kappa_3' \cos^2 \beta \right] \right\} /(1+R^2) 
\nonumber\\
&& +NR^2 \left[ \kappa_2'' - 2\kappa_3'' \right] 
\cos \gamma \sin \beta \cos \beta /(1+R^2) ~,
\nonumber\\
\eta_3 &=& -N \left\{ \kappa_1 + R^2 
\left[ \kappa_2 
\left( \cos^2 \gamma \sin^2(2\beta)+\cos^2(2\beta) \right) 
+ \kappa_3 + \kappa_4 \sin^2 \gamma \sin^2(2\beta) \right] 
\right\} /(1+R^2)
\nonumber\\
&& -N \left\{ -\kappa_1' \cos(2\beta) + R^2 \left[ \kappa_2' 
\left( -\cos^2(2\beta)+\cos^2 \gamma \sin^2(2\beta) \right) 
-2\kappa_3' \cos(2\beta) \right] \right\} /(1+R^2) 
\nonumber\\
&& -N \left\{ \kappa_1'' + R^2 \left[ -\kappa_2'' \cos(2\beta) 
+ 2\kappa_3'' \right] \right\} \cos \gamma \sin(2\beta)/(1+R^2) ~. 
\ea
The only nonvanishing coupling terms are $\eta_{uv}$, $\eta_{uw}$, 
$\eta_{vw}$, $\eta_{12}$, $\eta_{3v}$ and $\eta_{3w}$ 
\ba 
\eta_{uv} &=& 2N \left\{ \kappa_1' \sin (2\beta) 
+ \kappa_1'' \cos \gamma \cos(2\beta) \right\} /\sqrt{1+R^2} ~, 
\nonumber\\
\eta_{uw} &=& -2N \kappa_1'' \sin \gamma /\sqrt{1+R^2} ~, 
\nonumber\\
\eta_{vw} &=& -NR^2 \left[ \kappa_2-\kappa_4 \right] 
\sin(2\gamma) \cos(2\beta) /(1+R^2)
\nonumber\\
&& -NR^2 \kappa_2' \sin(2\gamma) \cos(2\beta) /(1+R^2)
\nonumber\\
&& -NR^2 \kappa_2'' \sin \gamma \sin(2\beta) /(1+R^2) ~, 
\ea
and 
\ba
\eta_{12} &=& -NR^2 \kappa_3 \cos \gamma \sin(2\beta) /(1+R^2)
\nonumber\\
&& -2NR^2 \kappa_2' \cos \gamma \sin(2\beta) /(1+R^2) 
\nonumber\\
&& -N \left\{ 2\kappa_1'' - R^2 \left[ \kappa_2'' \cos(2\beta) 
+ 2\kappa_3'' \right] \right\} /(1+R^2) ~,
\nonumber\\
\eta_{3v} &=& -NR^2 \left[ -\kappa_2+\kappa_4 \right] 
\sin(2\gamma) \sin(2\beta) /(1+R^2)
\nonumber\\
&& +NR^2 \kappa_2' \sin(2\gamma) \sin(2\beta) /(1+R^2)
\nonumber\\
&& +N \left\{ 2\kappa_1'' + R^2 \left[ -\kappa_2'' \cos(2\beta) 
+ 2\kappa_3'' \right] \right\} \sin \gamma /(1+R^2) ~,
\nonumber\\
\eta_{3w} &=& -NR^2 \left[ \kappa_2-\kappa_4 \right] 
\sin(4\beta) \sin^2 \gamma /(1+R^2) 
\nonumber\\
&& +N \left\{ 2\kappa_1' + R^2 
\left[ 2\kappa_2' \cos(2\beta)(1+\cos^2 \gamma) 
+ 2\kappa_3' \right] \right\} \sin(2\beta) /(1+R^2) 
\nonumber\\
&& +N \left\{ 2\kappa_1'' \cos(2\beta) 
+ R^2 \left[ -\kappa_2'' \cos(4\beta) 
+ 2\kappa_3'' \cos(2\beta) \right] \right\} \cos \gamma /(1+R^2) ~.
\ea

\subsection{Nonlinear XY$_2$ configurations}

For nonlinear XY$_2$ configurations the relevant point group 
is $C_{2v}$ \cite{Herzberg}. If we label the identical objects 
by 1 and 2, the Jacobi coordinates of Eq.~(\ref{jacobi}) with 
$m_1=m_2$ become perpendicular. The corresponding equilibrium shape is 
\ba
\overline{r} \;=\; R \;>\; 0 ~, 
\hspace{1cm} \overline{\chi} \;=\; \beta ~, 
\hspace{1cm} \overline{\theta} \;=\; \gamma \;=\; \pi/2 ~. 
\label{xy2}
\ea 
The permutation symmetry associated with the interchange of the 
two identical objects $S_2$ (isomorphic to $D_2$) is determined 
by the transposition $P(12)$. This excludes from the Hamiltonian 
all terms that change the number of $p_{\rho}$ bosons by an odd number. 

Accordingly, the intrinsic Hamiltonian for nonlinear XY$_2$ 
configurations is given by Eqs.~(\ref{hint1}) and (\ref{hint2}) 
with the shape variables of Eq.~(\ref{xy2}), and 
\ba
E \;=\; F \;=\; 0 ~.
\ea
The normal modes are obtained as before by carrying out a 
mean-field analysis of the intrinsic Hamiltonian 
\ba
^{B}H_{\mbox{int}} 
&=& \epsilon_u \, b^{\dagger}_{u} b_{u}
  + \epsilon_v \, b^{\dagger}_{v} b_{v}
  + \epsilon_w \, b^{\dagger}_{w} b_{w}
  + \epsilon_{uv} ( b^{\dagger}_{u} b_{v} + b^{\dagger}_{v} b_{u} ) ~.
\ea
In this case, the two radial modes, $u$ and $v$, are decoupled from 
the angular mode $w$. The normal radial modes 
are obtained by diagonalizing a $2 \times 2$ matrix, and correspond 
to the symmetric and antisymmetric stretching modes. This 
diagonalization is the algebraic analogue of the orthogonal 
transformation of Delves coordinates ({\it i.e} mass-scaled 
Jacobi coordinates) to the normal coordinates in the small 
displacement limit \cite{bowman}.

Similarly, the collective Hamiltonian is given by 
Eqs.~(\ref{hcol1}) and (\ref{hcol2}) with 
\ba
H^{(E_{\rho})}_{\mbox{c}} \;=\; 0 ~.
\ea
Its Bogoliubov image has the form
\ba
^{B}H_{\mbox{c}} 
&=&  \eta_u \, ( b^{\dagger}_u - b_u )^2 
   + \eta_v \, ( b^{\dagger}_v - b_v )^2 
   + \eta_w \, ( b^{\dagger}_w - b_w )^2 
\nonumber\\
&& + \eta_1 \, ( b^{\dagger}_1 - b_1 )^2 
   + \eta_2 \, ( b^{\dagger}_2 - b_2 )^2 
   + \eta_3 \, ( b^{\dagger}_3 - b_3 )^2 
\nonumber\\
&& + \eta_{uv} \, ( b^{\dagger}_u - b_u ) \, ( b^{\dagger}_v - b_v )
   + \eta_{3w} \, ( b^{\dagger}_3 - b_3 ) \, ( b^{\dagger}_w - b_w ) ~.
\ea
In addition to the $\eta_i$ terms, 
the only coupling terms present are $\eta_{uv}$ and $\eta_{3w}$. 
In comparison with the nonlinear XYZ configuration, the coupling 
terms $\eta_{uw}$, $\eta_{vw}$, $\eta_{12}$ and  
$\eta_{3v}$ are missing. This is a consequence of the different 
point group symmetries.

\section{Oblate symmetric top}
 
Nonlinear X$_3$ configurations have the equilibrium shape of an 
equilateral triangle, which is characterized by two orthogonal 
Jacobi coordinates of equal length. 
The geometric configuration has one 3-fold 
symmetry axis, three 2-fold symmetry axes perpendicular to it and one
reflection plane. The corresponding point group is $D_{3h}$ whose
classification is equivalent to that of its subgroup $D_3$ 
(which is isomorphic to the permutation group of three identical 
objects $S_3$) and parity. 
In comparison with nonlinear XY$_2$ configurations, for three 
identical objects the Jacobi coordinates not only are perpendicular, 
but also have equal lengths. The equilibrium shape is 
then characterized by 
\ba
\overline{r} \;=\; R \;>\; 0 ~, 
\hspace{1cm} \overline{\chi} \;=\; \beta \;=\; \pi/4 ~, 
\hspace{1cm} \overline{\theta} \;=\; \gamma \;=\; \pi/2 ~. 
\label{x3}
\ea 
The permutation symmetry associated with the interchange of the 
three identical objects $S_3$ (isomorphic to $D_3$) is determined 
by the transposition $P(12)$ and the cyclic permutation $P(123)$. 
This excludes from the Hamiltonian all terms that are not scalar with 
respect to $S_3$. The $S_3$ invariant Hamiltonian (or mass operator) 
is discussed in more detail in \cite{BIL,BDL}. Here we briefly 
present the results that are relevant to the present article. 

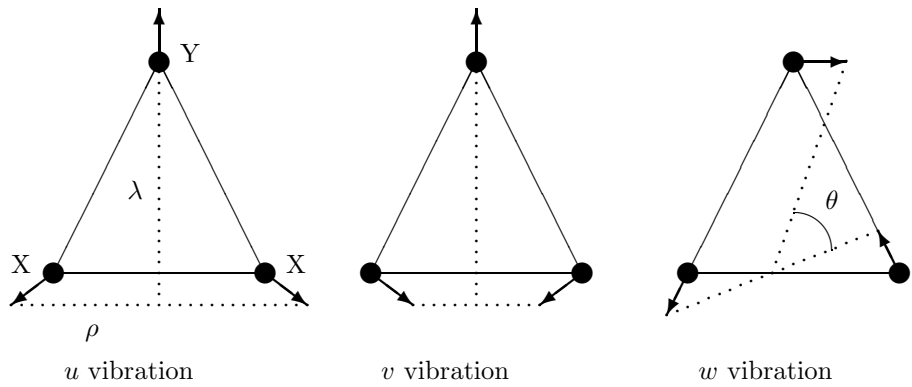
\begin{figure}
\centering
\setlength{\unitlength}{0.8pt}
\begin{picture}(450,175)(0,0)
\thinlines
\put ( 30,  0) {$u$ vibration}
\put ( 25, 50) {\circle*{10}}
\put (125, 50) {\circle*{10}}
\put ( 75,150) {\circle*{10}}
\put (  5, 50) {X}
\put ( 85,150) {Y}
\put (135, 50) {X}
\put ( 25, 50) {\line ( 1,0){100}}
\put ( 25, 50) {\line ( 1,2){ 50}}
\put (125, 50) {\line (-1,2){ 50}}
\multiput (  5, 35)( 5, 0){29}{\circle*{0.1}}
\multiput ( 75,150)( 0,-5){23}{\circle*{0.1}}
\put ( 40, 20) {$\rho$}
\put ( 60, 85) {$\lambda$}
\thicklines
\put ( 75,150) {\vector( 0, 1){25}}
\put ( 25, 50) {\vector(-4,-3){20}}
\put (125, 50) {\vector( 4,-3){20}}
\thinlines
\put (180,  0) {$v$ vibration}
\put (175, 50) {\circle*{10}}
\put (275, 50) {\circle*{10}}
\put (225,150) {\circle*{10}}
\put (175, 50) {\line ( 1,0){100}}
\put (175, 50) {\line ( 1,2){ 50}}
\put (275, 50) {\line (-1,2){ 50}}
\multiput (195, 35)( 5, 0){13}{\circle*{0.1}}
\multiput (225,150)( 0,-5){23}{\circle*{0.1}}
\thicklines
\put (225,150) {\vector( 0, 1){25}}
\put (175, 50) {\vector( 4,-3){20}}
\put (275, 50) {\vector(-4,-3){20}}
\thinlines
\put (330,  0) {$w$ vibration}
\put (325, 50) {\circle*{10}}
\put (425, 50) {\circle*{10}}
\put (375,150) {\circle*{10}}
\put (325, 50) {\line ( 1,0){100}}
\put (325, 50) {\line ( 1,2){ 50}}
\put (425, 50) {\line (-1,2){ 50}}
\put (390, 80) {$\theta$}
\multiput (315, 30)(   5,2){21}{\circle*{0.1}}
\multiput (365, 50)(1.75,5){21}{\circle*{0.1}}
\put (375, 60.5) {\oval(36,36)[tr]}
\thicklines
\put (375,150) {\vector( 1, 0){25}}
\put (325, 50) {\vector(-1,-2){10}}
\put (425, 50) {\vector(-1, 2){10}}
\end{picture}
\caption[]{\small Schematic representation of the normal vibrations
of a nonlinear X$_3$ configuration. The Jacobi coordinates are indicated
by the dotted lines.   
\normalsize}
\label{fund}
\end{figure}

The intrinsic Hamiltonian for nonlinear X$_3$ 
configurations is given by Eqs.~(\ref{hint1}) and (\ref{hint2}) 
with the shape variables of Eq.~(\ref{x3}), and 
\ba
B \;=\; C ~, \hspace{1cm} D \;=\; E \;=\; F \;=\; 0 ~. 
\ea
As before, the normal modes are obtained by 
a mean-field analysis 
\ba
^{B}H_{\mbox{int}} 
&=& \epsilon_u \, b^{\dagger}_{u} b_{u} 
+ \epsilon_v \, ( b^{\dagger}_{v} b_{v} + b^{\dagger}_{w} b_{w} ) ~.
\ea
In this case, the two radial modes, $u$ and $v$, and the angular mode 
$w$ are decoupled, and correspond to the fundamental vibrations 
(see Fig.~\ref{fund}) of the configuration of Fig.~\ref{geometry}.
The first term represents the symmetric stretching 
mode $u$, whereas the second term corresponds to a degenerate doublet 
($\epsilon_v=\epsilon_w$) of the antisymmetric stretching mode $v$ and
the bending mode $w$. This is in agreement
with the point-group classification of the fundamental vibrations
of a symmetric-top $X_3$ configuration \cite{Herzberg}.

The collective Hamiltonian is given by Eqs.~(\ref{hcol1}) 
and (\ref{hcol2}) with 
\ba 
H^{(E_{\lambda})}_{\mbox{c}} \;=\; H^{(E_{\rho})}_{\mbox{c}} \;=\; 0 ~,
\ea
and its Bogoliubov image has the form 
\ba
^{B}H_{\mbox{c}}
&=& \eta_u \, ( b^{\dagger}_u - b_u )^2 
  + \eta_v \, [ ( b^{\dagger}_v - b_v )^2 + ( b^{\dagger}_w - b_w )^2 ] 
\nonumber\\
&&+ \eta_1 \, [ ( b^{\dagger}_1 - b_1 )^2 + ( b^{\dagger}_2 - b_2 )^2 ] 
  + \eta_3 \, ( b^{\dagger}_3 - b_3 )^2 ~.
\ea
The terms which depend on the vibrational bosons $b_i^{\dagger}$ 
($i=u,v,w$) cause a shift in the vibrational
frequencies, whereas the terms involving the rotational bosons
$b_i^{\dagger}$ ($i=1,2,3$) are responsible for in-band rotational
splitting. The equality of the moments of inertia about the $x$ and 
$y$ axis ($\eta_1=\eta_2$), and the equality of the coefficients 
$\eta_v=\eta_w$ reflect the axial symmetry of the underlying shape. 
In comparison with the nonlinear XYZ and XY$_2$ configurations 
all coupling terms $\eta_{ij}$ are missing.

The above analysis presents a clear interpretation of the various
interaction terms in the intrinsic and collective Hamiltonians.
The two terms in $H_{\mbox{int}}$ proportional to $A$ and $B=C$ 
contribute only to the eigenfrequencies of the vibrational modes.
The $\kappa_3$ and $\kappa_4$ terms in $H_{\mbox{c}}$ 
contribute only to the rotational modes, whereas the 
$\kappa_1$ and $\kappa_2$ terms both effect the vibrational and the
rotational modes, and hence represent vibration-rotation
couplings. The equality of the frequencies of the antisymmetric 
stretching and the bending modes, and the equality of the moments of 
inertia about the $x$ and $y$ axis, show that the above Hamiltonians 
correspond to an oblate symmetric top with the threefold symmetry 
axis along the $z$-axis. In \cite{BDL} it was shown that $U(7)$ 
provides a complete classification of the vibrational and rotational 
excitations of the oblate symmetric top. 

\section{Summary and conclusions}

We have presented an algebraic treatment of three-body problems.
The relative motion of the three-body system is treated by the method
of bosonic quantization, which for the two relative vectors
gives rise to a $U(7)$ spectrum generating algebra. The model
space is spanned by the symmetric irreducible representation
$[N]$ of $U(7)$. In particular, we have considered nonlinear XYZ, 
XY$_2$ and X$_3$ configurations. The $U(7)$ interaction terms have 
been interpreted geometrically by means of coherent states and a
normal mode analysis.  
We have studied the effect of the permutation symmetry among the 
identical particles on the coupling between the vibrational and 
rotational modes. For XYZ and XY$_2$ the geometry is that of an 
asymmetric top, whereas for $X_3$ that of an oblate top. 
The ensuing algebraic treatment of the oblate top has found useful
applications both in molecular physics (X$_3$ molecules
\cite{BDL}) and hadronic physics (nonstrange $qqq$ baryons
\cite{BIL,emff}). 

In conclusion, we have shown that $U(7)$ which was previously 
introduced to describe X$_3$ configurations \cite{BIL,BDL}, 
also provides a spectrum generating algebra for XY$_2$ and XYZ 
configurations. The latter are important to describe nonlinear molecules
and strange baryons. In this respect, $U(7)$ provides a unified 
treatment of vibrational and rotational excitations of both 
symmetric and asymmetric tops. 

\section*{Acknowledgements}

It is a pleasure to dedicate this paper to the 60th birthday
of John Tjon and his many important contributions to physics.
This work is supported in part by DGAPA-UNAM under project IN101997,
and by grant No. 94-00059 from the United States-Israel Binational
Science Foundation.

\clearpage

\clearpage

\begin{thebibliography}{99}

\bibitem{fbs}
See {\it e.g.} Proceedings of the Thirteenth International
Conference on Few Body Problem in Physics, Adelaide,
S.A., Australia, Eds. I.R. Afnan and R.T. Cahill,
Nucl. Phys. A {\bf 543} (1992);
Proceedings of the 15th European Conference on Few-Body Problems
in Physics, Pe\~niscola, Spain, Ed. R. Guardiola,
Few-Body Systems, Supplement 8, Springer Verlag, 1996.

\bibitem{ibm}
A. Arima and F. Iachello,
Phys. Rev. Lett. {\bf 35}, 1069 (1975); for a review see
F. Iachello and A. Arima, `The interacting boson model',
Cambridge University Press, 1987.

\bibitem{vibron}
F. Iachello,
Chem. Phys. Lett. {\bf 78}, 581 (1981); for a review see
F. Iachello and R.D. Levine, `Algebraic theory of molecules',
Oxford University Press, 1995.

\bibitem{FI}
F. Iachello,
Nucl. Phys. A {\bf 560}, 23 (1993).

\bibitem{BIL}
R. Bijker, F. Iachello and A. Leviatan,
Ann. Phys. (N.Y.) {\bf 236}, 69 (1994).

\bibitem{BDL}
R. Bijker, A.E.L. Dieperink, A. Leviatan,
Phys. Rev. A {\bf 52}, 2786 (1995).

\bibitem{KM}
P. Kramer and M. Moshinsky,
Nucl. Phys. {\bf 82}, 241 (1966).

\bibitem{Mitra}
A.N. Mitra, A. Sharma and B. Mitra-Sodermark, 
Few-Body Systems {\bf 19}, 145 (1995).

\bibitem{cs}
For a general review on coherent states see {\it e.g.}: 
W.M. Zhang, D.H. Feng and R. Gilmore, 
Rev. Mod. Phys. {\bf 62}, 867 (1990);
For an application to the interacting boson model see {\it e.g.}: 
A. Bohr and B.R. Mottelson, 
Physica Scripta {\bf 22}, 468 (1980); 
J.N. Ginocchio and M.W. Kirson, 
Phys. Rev. Lett. {\bf 44}, 1744 (1980);  
A.E.L. Dieperink, O. Scholten and F. Iachello, 
Phys. Rev. Lett. {\bf 44}, 1747 (1980). 

\bibitem{hyper}
See {\it e.g.} L.M. Delves, Nucl. Phys. {\bf 9}, 391 (1959);
F.T. Smith, Phys. Rev. {\bf 120}, 1058 (1960);
J.L. Ballot and M. Fabre de la Ripelle, 
Ann. Phys. (N.Y.) {\bf 127}, 62 (1980).

\bibitem{KL}
M.W. Kirson and A. Leviatan,
Phys. Rev. Lett. {\bf 55}, 2846 (1985).

\bibitem{Herzberg}
G. Herzberg, `The spectra and structures of simple free radicals',
Dover, 1971.

\bibitem{bowman}
J.M. Bowman, J. Zuniga and A. Wierzbicki,
J. Chem. Phys. {\bf 90}, 2708 (1989).

\bibitem{emff}
R. Bijker, F. Iachello and A. Leviatan,
Phys. Rev. C {\bf 54}, 1935 (1996); 
R. Bijker, F. Iachello and A. Leviatan, 
Phys. Rev. D {\bf 55}, 2862 (1997).

\end{thebibliography}
\end{document}